\documentclass[journal=jacsat,manuscript=article]{achemso}

\usepackage[version=3]{mhchem} 
\usepackage{amsmath,amssymb}
\usepackage{xcolor}

\author{H. J. Muhren}

\author{Paul van der Schoot}

\email{p.vanderschoot@phys.tue.nl}

\affiliation[TUE]
{Soft Matter and Biological Physics, Department of Applied Physics and Science Education, Eindhoven University of Technology, Postbus 513, 5600 MB Eindhoven, The Netherlands}

\title[Donnan theory viruses]
  {Electrostatic theory of the acidity of the solution in the lumina of viruses and virus-like particles}


\begin{document}

\begin{tocentry}
\begin{center}
\includegraphics[scale=0.11]{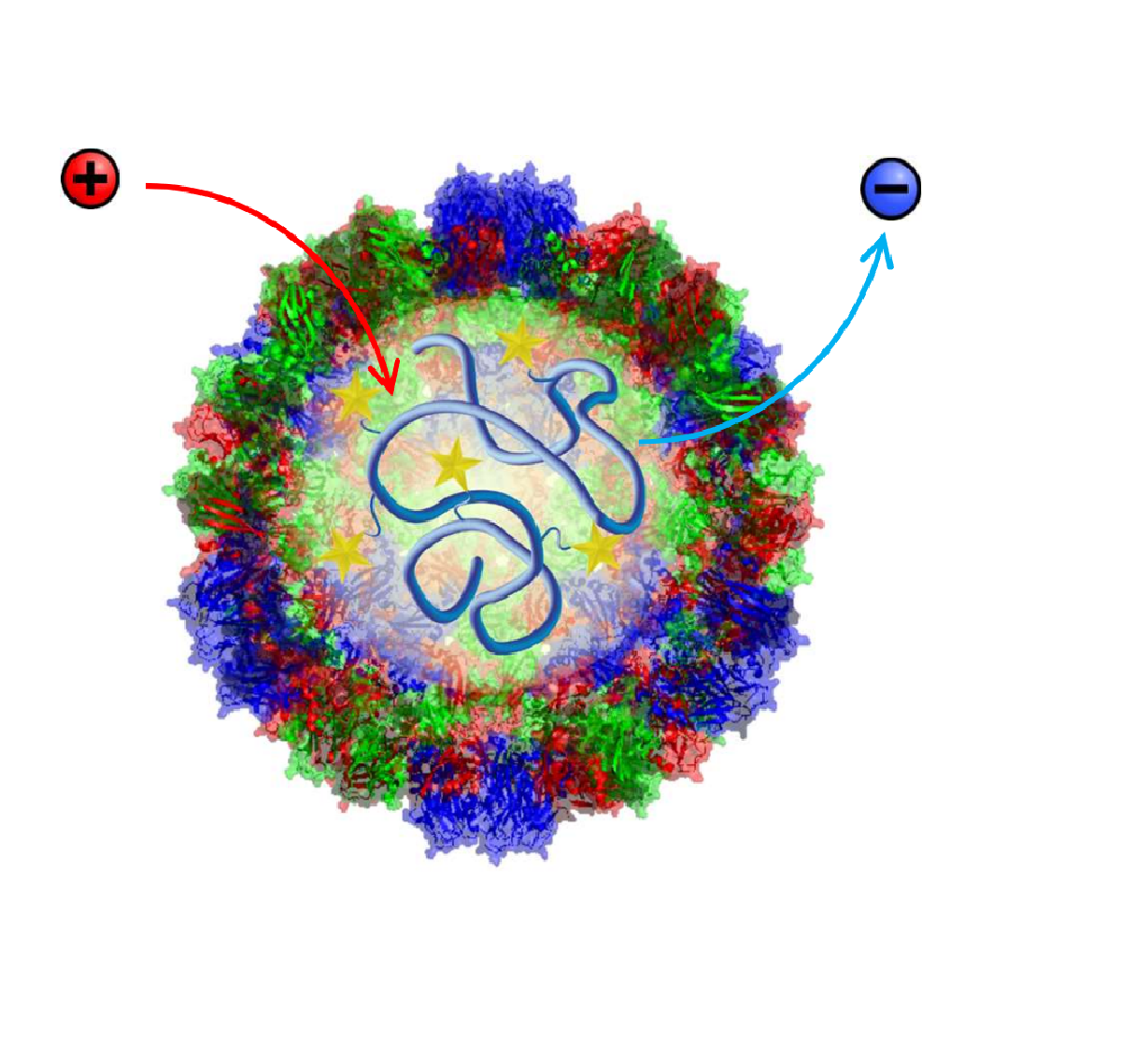}
\end{center}

Poisson-Boltzmann theory confirms that Donnan theory accurately describes the pH difference between the lumen and the bulk solution of viruses, virus-like particles and other protein shells.
\end{tocentry}

\begin{abstract}
Recently, Maassen \textit{et al.}\ measured an appreciable pH difference between the bulk solution and the solution in the lumen of virus-like particles, self-assembled in an aqueous buffer solution containing the coat proteins of a simple plant virus and polyanions. [Maassen, S. J.; \textit{et al.}\ Small \textbf{2018}, \textit{14}, 1802081] They attribute this to the Donnan effect, caused by an imbalance between the number of negative charges on the encapsulated polyelectrolyte molecules and the number of positive charges on the RNA binding domains of the coat proteins that make up the virus shell or capsid. 
By applying Poisson-Boltzmann theory, we confirm this conclusion and show that simple Donnan theory is accurate even for the smallest of viruses and virus-like particles. This, in part, is due to the additional screening caused by the presence of a large number of immobile charges in the cavity of the shell. The presence of a net charge on the outer surface of the capsid we find in practice to not have a large effect on the pH shift. Hence, Donnan theory can indeed be applied to connect the local pH and the amount of encapsulated material. The large shifts up to a full pH unit that we predict must have consequences for applications of virus capsids as nanocontainers in  bionanotechnology and artificial cell organelles.
\end{abstract}

\section{Introduction}
The co-assembly of the coat proteins of many simple viruses with their single-stranded (ss) RNA seems to be primarily driven by electrostatic interactions between the negatively charged genome and the positively charged, disordered RNA binding domains on the coat proteins. \cite{McPherson2005,Roos2020,Gelbart2016,ChengKao2012} It is no surprise, then, that under appropriate solution conditions virus coat proteins spontaneously encapsulate not only homologous but also heterologous ssRNAs, synthetic polyanions, surface-functionalized nanoparticles and so on.\cite{Bancroft1969,Young1998,Dragnea2007,Gelbart2008,Cornelissen2010,Gelbart2011,Cornelissen2019} Interestingly, the co-assembly does not necessarily produce virus-like particles of the same size or T number or even the same shape as that of the native virus. Size and shape selection seems to be controlled on the one hand by mass action and hence stoichiometry, and on the other by a compromise between the interaction between the coat proteins and that between the coat proteins and cargo.\cite{Gelbart2008,vanderSchoot2009,vanderSchoot2015}
Virus coat proteins, such as that of Brome Mosaic Virus, Cowpea Chlorotic Mosaic Virus and Hepatitis B Virus, can actually be made to self-assemble into shells in the absence of a negatively charged cargo by changing the acidity and/or salinity of the solution, removing the RNA binding domain of the coat proteins or chemically modifying this domain.\cite{Zlotnick2002,Lavelle2009,Timmermans2022} The spontaneous assembly is in that case driven by hydrophobic and other types of attractive interaction, involving, \textit{e.g.}, ionic and hydrogen bonds.\cite{KegelSchoot2004,Podgornik2020} 

Because virus coat proteins are able to encapsulate molecular cargo, either spontaneously from solution or by attaching it chemically or physically to, e.g., the RNA binding domain of the protein, there is a significant interest in utilising this property for application purposes in targeted drug delivery, tomography, controlled catalysis and metamaterials.\cite{Nolte2012,Steinmetz2015,Steinmetz2016,Sainsbury2022} 
Recently, virus-based artificial organelles have been suggested as a viable route to be used in living cells for therapeutic purposes, restoring or even adding cellular activity.\cite{VanHest2021} These artificial organelles contain catalytically active particles such as enzymes. The reason that encapsulated catalytic particles can actively process substrates present in the solution, is that the capsid shell is permeable to these molecules but only if they are not too large in size. If too large, substrates cannot diffuse through the  pores in the shell that include ones with a diameter as large as a few nanometers.\cite{Douglas2021} The semi-permeability of capsids, in particular that of plant viruses, has been known for some time.\cite{Bancroft1984}  

Catalytic activity, in general, is strongly pH and ionic strength dependent, suggesting that the control of the physico-chemical conditions of the solution in the lumen of the virus capsid or any other type of proteinaceous shell used for the same purpose must be of paramount importance.\cite{Lovell2022} Interestingly, so far this issue seems to have met with relatively little attention in the artificial organelle community,\cite{VanHest2021} even though recent experiments by Maassen \textit{et al.}\ demonstrate that the pH in the lumen of a virus-like particle can be much lower than that of the bulk solution.\cite{Cornelissen2018} It is interesting to mention that the lumina of carboxysomes, which are proteinaceous shells that act as organelles in bacteria, also seem to have a lower pH than the bulk solution both \textit{in vitro} and \textit{in vivo}.\cite{Liu2022} The reason that the acidity and ionic composition of the lumen of a protein shell can be much different than that of the bulk solution is the presence of a net \textit{immobile} charge in that lumen, that is, a net charge not associated with translationally mobile species that can freely diffuse in and out of the shell, and the result of what is commonly known as the Donnan effect.\cite{Vrij2011} This effect has long been known to cause pH gradients across lipid membranes in the context of vesicles.\cite{Dubinsky1985}

In the experiments of Maassen \textit{et al.},\cite{Cornelissen2018} the coat protein of cowpea chlorotic mottle virus (CCMV) and a 7.5 kDa random copolymer of styrene sulfonate and  pH-sensitive fluorescent fluorescein methacrylate monomers spontaneously assemble in virus-like particles under a wide range of solutions pHs. The authors use the ratio of two excitation peaks to obtain the relation between the pH in free solution and that in the lumen. For pHs spanning values between 6.0 and 8.0, the pH inside the virus-like particle turns out to be about 0.4 pH units \textit{lower} than that outside of it, that is, in the bulk solution. A simple Donnan theory put forward by the authors, presuming a uniform distribution of immobile charges in the lumen and ignoring both the presence of the protein shell holding the polyanionic cargo and the presence of a net charge on the outer surface of it, actually explains this observation. 
In the present context, Donnan theory presumes (i) \textit{local} charge neutrality, meaning in this case that both the capsid lumen and the bulk solution outside the virus are electroneutral, and (ii) equal chemical potentials of the mobile ionic species in- and outside of the shells.\cite{Donnan1911,Donnan1924} 
Typically, the well-known expressions for the electrochemical potentials of the charged mobile ionic species valid in dilute solution are used for the latter. Standard chemical potentials and the effects of non-ideality of the solution, insofar that these can be described by activity coefficients, can be absorbed in the Donnan potential, which actually acts as a Lagrange multiplier enforcing local charge neutrality. This implies that the theory should be valid even if the ionic solution does not behave ideally.\cite{Cornelissen2018}

According to the theory, the dimensionless Donnan potential $\phi$ in the capsid lumen, defined as the Donnan potential energy scaled to the thermal energy, obeys the simple relation\cite{Samelson1952,Lifson1957}
\begin{equation}
   \phi = \sinh^{-1} \gamma,
\end{equation}
where the quantity $\gamma \equiv \Delta \rho / 2 \rho_\mathrm{s}$ is defined in terms of the difference $\Delta \rho = \rho_+^\mathrm{i}-\rho_-^\mathrm{i}$ of the number densities of positive and negative immobile charges $\rho_\pm^\mathrm{i}$, averaged over the volume of the lumen, and $2 \rho_\mathrm{s}$ the overall density of mobile ionic species in the bulk solution (tacitly presumed to be monovalent). This quantity is dominated by the concentrations of added salt and buffer of the assembly mixture, and explains why in practise the Donnan potential and also the pH differential across the protein shell is virtually independent of the acidity of the solution,\cite{Cornelissen2018,Liu2022} unless some form of charge regulation takes place involving weakly ionic moieties on the cargo or the coat proteins themselves.\cite{Mohwald2007,Cornelissen2018} 

The pH shift, defined as the difference $\Delta \mathrm{pH} = \mathrm{pH_{in}} - \mathrm{pH_{out}}$ between the solution pH$_\mathrm{in}$ inside and pH$_\mathrm{out}$ outside of the particle, is proportional to the Donnan potential,
\begin{equation}
    \Delta \mathrm{pH} = \mathrm{pH_{in}} - \mathrm{pH_{out}} = \frac{\phi}{\ln 10},
\end{equation}
which follows from the Boltzmann distribution that connects the concentrations of the mobile ionic species in- and outside of the shell. With a value of $\Delta \mathrm{pH} \simeq -0.4$ obtained from the experiments, eqs.\ (1) an (2) suggest a value of $\gamma \simeq -1.1$. The ionic strength of the mobile ionic species present in the lumen is as a result of this almost fifty per cent larger than that in the bulk solution, which is physiological and equal to  0.15 M.\cite{Cornelissen2018} Computer simulations confirm that the mobile ionic content of the capsid lumen must be very different from that in the bulk solution.\cite{Linse2007}

A negative value of $\gamma$ implies that a larger number of negative polymer charges have been encapsulated than necessary to compensate for the positive charges on the RNA binding domains. This phenomenon is usually referred to as charge reversal or overcharging, and $\gamma$ can therefore be seen as a measure for the degree of over- or undercharging.\cite{Bruinsma2005,Muthukumar2006,Shklovskii2008,Wang2011} Taking as a radius for the lumen a range of $4 - 5$ nm, an ionic strength of 0.15 M, and 600 positive charges on the RNA binding domains, we obtain for the \textit{relative} overcharging $-\Delta \rho/\rho_+^\mathrm{i}$ a value corresponding to $8 - 15\%$. This translates to between 10 and 18 encapsulated chains.\cite{Cornelissen2018} Conversely, we can use the fact that a large number of $T=3$ viruses with a genome smaller than about 6000 nt seem to have an average degree of overcharging of $+60\%$,\cite{Podgornik2018} and convert this number into a pH shift. According to Donnan theory, this degree of overcharging translates to a pH shift of $-1$, if we take the average of 10 or so positive charges per RNA binding domain of those viruses.\cite{Podgornik2018} 
Overall, relative degrees of overcharging ranging from $-100\%$ for empty capsid shells due to the absence of any charge compensation by immobile cargo because in that case $\rho_-^\mathrm{i} = 0$, to $+800\%$ for encapsulated high-molecular weight poly(styrene sulfonate) have been reported.\cite{Podgornik2020} This then suggests that, if we take eqs.\ (1) and (2) at face value, there must be a large spread in pH shift in virus-like particles ranging in value from about $-1$ and $+1$ depending on the amount of encapsulated cargo, ionic strength and so on. 

Of course, these estimates hold only if Donnan theory, which presumes a spatially uniform electrochemical potential in the lumen of the protein shell, actually applies on the scale of $T=1$ and $T=3$ viruses. Such small viruses measure between twenty to thirty nanometers in outer diameter, whilst their lumina are obviously even smaller than that, and, say, ten to twenty nanometers wide.\cite{Podgornik2012} 
Typical electrostatic screening lengths are around one nanometer and any encapsulated polyanion should perhaps be expected to be concentrated primarily in the region where the RNA binding domains are located, which for CCMV for instance is estimated to be about three or four nanometers wide.\cite{Muthukumar2006,ChengKao2012,vanderSchoot2015} Theoretically, the effect seems to be smaller for linear polyanions than for ssRNA-like randomly branched ones, and for $T=1$ particles this layer should extend almost to the centre of the cavity of the capsid.\cite{vanderSchoot2013,Tresset2021} 
One might attempt to construct a non-uniform Donnan theory, in a similar vein as was done by Odijk and Slok for densely packed double-stranded DNA in bacteriophages\cite{Odijk2003},  and by Philipse in the somewhat different context of charged colloids in a gravitational field.\cite{Philipse2004} In the present context of over- or undercharged virus-like particles such an enterprise would defeat our purpose, which is to obtain estimates from a simple theory that requires as few input parameters as possible, and that does not require taking recourse to numerical methods.

Even if the distribution of the immobile charges in the lumen is more or less uniform, as we expect it to be for $T=1$-sized shells, we nevertheless think it is prudent to investigate under what conditions simple Donnan theory applies, and how accurate it actually is given that the mobile charges tend to \textit{not} be uniformly distributed.\cite{Lifson1957} We note also that within Donnan theory the presence of a net charge on the outer surface of the shell, which can be net positive or net negative depending on the type of virus and pH of the solution,\cite{Podgornik2012} does \textit{not} affect the predicted pH shift. The reason is that for typical concentrations of coat protein up to a hundred $\mu$M in \textit{in vitro} experiments and typical numbers of surface charges of perhaps a few hundred, the corresponding concentration of surface charges should remain much lower than the ionic strength of the buffer solution that typically is (near) physiological. 

By applying Poisson-Boltzmann theory to model virus-like particles, we are able confirm that predictions for pH shifts obtained from Donnan theory are quite accurate also for the smallest, that is, the $T=1$-sized particles, even for ionic strengths much below physiological. The reason is that in practice the overall concentration of immobile ionic species in the capsid lumen is very much larger than the ionic strength of the buffer solution, and because of this also that of the mobile ionic species. This gives rise to a smaller effective screening length in the capsid  lumen, which may in fact also be inferred from earlier work on the electrostatics of soft particles.\cite{Lifson1957,Duval2005,Ohshima2008} The impact of a surface charge is relatively minor, in part caused by the presence of the capsid shell. Interestingly, a net charge on the shell that has the same sign as the net immobile charge in the lumen potentially increases the accuracy of Donnan theory.

The remainder of this paper is organised as follows. First, in the section Theory, we formulate a Poisson-Boltzmann theory for a spherical charge distribution inside of a shell that has a net positive or negative surface charge on its outer surface. Following Lifson and others,\cite{Samelson1952,Lifson1957,Ohshima2008,Duval2022} we write the electrical potential in the cavity of the shell as the sum of a uniform and a position dependent part that we treat as a pertubation, where we identify the uniform part as the Donnan potential. We solve the equations analytically for the position-dependent potential (i) upto linear order in the pertubation in the cavity, (ii) exactly in the shell and (iii) at the level of a Debye-H\"{u}ckel approximation in the outer region. Next, in the section Results and Discussion, we analyse our findings, the most important one being that under a wide range of conditions the correction to the pH shift obtained from Donnan theory, caused either by non-uniformity of the electrical potential or the presence of surface charges on the shell, is minor. 
Finally, we provide a summary of our findings in the section Discussion and Conclusions, and discuss how both Donnan theory and Poisson-Boltzmann theory can be amended to account for the effects of charge regulation and explain why in the experiments of Maassen and collaborators the pH in the capsid lumen becomes constant below a pH of 6.\cite{Cornelissen2018}
 
\section{Theory}
To set up the Poisson-Boltzmann theory, we need to construct a model. In our model, we presume that (i) the geometry of the problem obeys spherical symmetry and (ii) all charges associated with the RNA binding domains of the coat proteins and cargo (the ``immobile'' charges) are uniformly distributed in a spherical volume of radius $R_1>0$. See also Figure 1. So, the number densities $\rho_\pm^\mathrm{i}(r)=\rho_\pm^\mathrm{i} H(R_1-r)$ of positive and negative immobile charges are step functions of the radial co-ordinate $r\in[0,\infty)$, with $\rho_\pm^\mathrm{i}$ (as before) their mean densities in the cavity and $H(R_1-r)=1$ for $r<R_1$ and $H(R_1-r)=0$ for $r\geq R_1$ the usual Heaviside step function. The number density associated with the net immobile space charge is defined as $\Delta \rho (r) \equiv \rho_+^\mathrm{i}(r)-\rho_-^\mathrm{i}(r) = (\rho_+^\mathrm{i}-\rho_-^\mathrm{i})H(R_1-r) \equiv \Delta \rho H(R_1-r)$. The quantity $\Delta \rho \equiv \rho_+^\mathrm{i}-\rho_-^\mathrm{i}$ is (as before) the net average number density of immobile charge in the lumen of the particle. It can take both positive and negative values, or be equal to zero.  

Surrounding the spherical volume is a shell of radius $R_2\geq R_1$. This shell is permeable to mobile ions due to the presence of holes that we do not explicitly model. So, we presume the shell to not contain any mobile or immobile charges, again for reasons of simplicity. On the outer surface of the shell, for $r=R_2$, we envisage the presence of a net number surface density of immobile charges $\Delta \sigma$, noting that this quantity can take positive and negative values or be equal to zero, depending on the type of virus and acidity of the solution. Our model resembles that of Duval and co-workers describing the electrokinetics of the bacteriophage MS2 and end-carboxylated dendrimers.\cite{Duval2008,Duval2015} We also note that any transient exposure of the RNA binding domains and/or cargo to the outside of the capsid shell,\cite{Bothner2015} e.g., via the pores, only renormalizes the charge densities of the lumen and the surface.

\begin{figure}[ht!]
\centering
\includegraphics[scale=0.6]{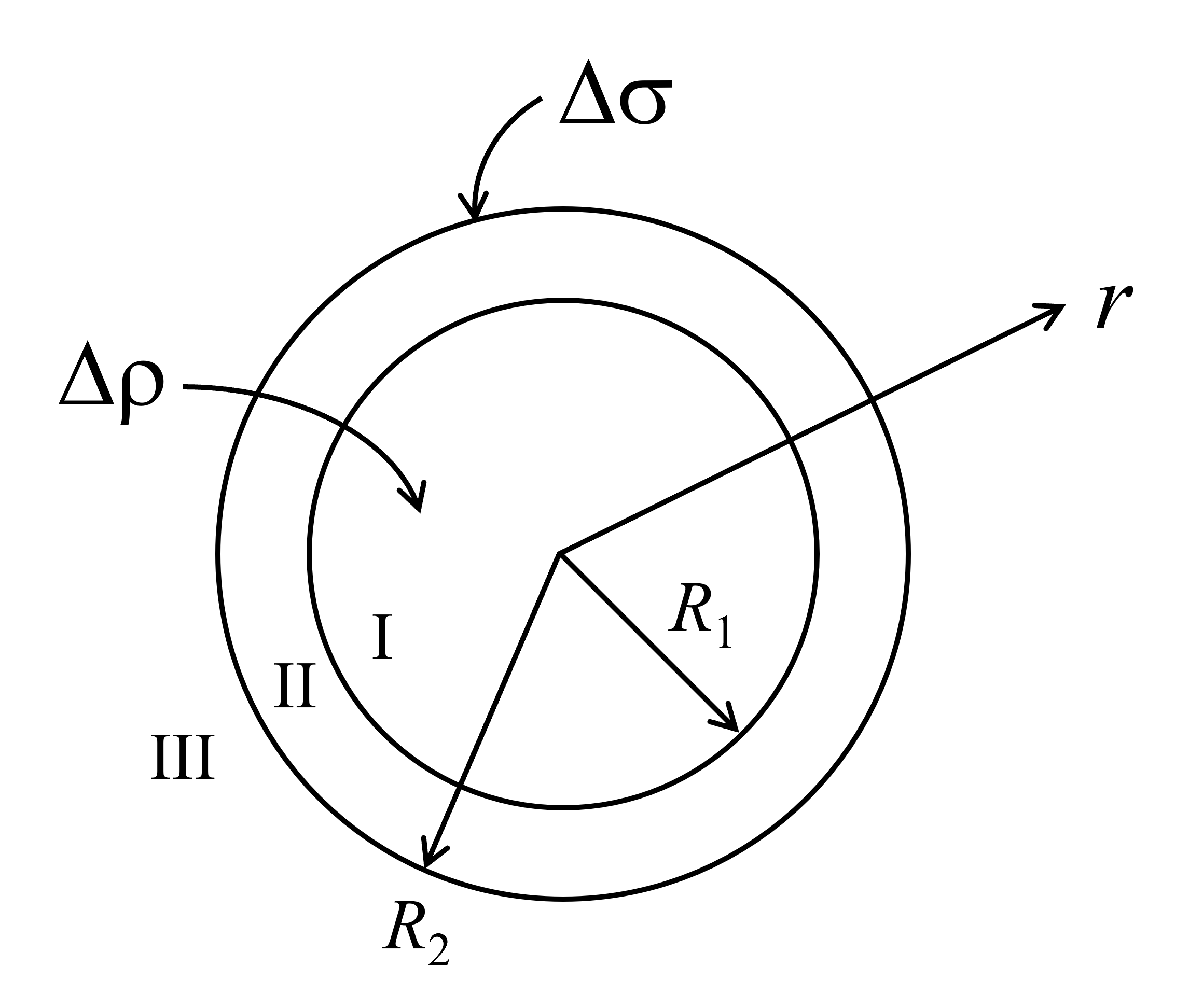}
\caption{Model of the virus or virus-like particle. Region I is the interior of the protein shell or capsid, for radial distances $r<R_1$. In this region, there is a net immobile space charge of number density $\Delta \rho$. Region II represents the capsid, with $R_1\leq r\leq R_2$. The outer surface of the capsid is characterised by the presence of a net immobile surface charge of surface number density $\Delta \sigma$. Region III, defined by $r>R_2$, represents the solution outside of the particle. For $r\gg R_2$ the solution becomes the bulk solution. See also the main text.}
\end{figure}

Using the known $\mathrm{pK_a}$s of basic and acidic residues of surface-exposed amino acids, Lo\v{s}dorfer Bo\v{z}i\v{c} and collaborators predict that for a large number of viruses $-0.3 \lesssim \Delta \sigma \lesssim +0.1$ nm$^{-2}$ at neutral pH albeit that for the vast majority $-0.2 \lesssim \Delta \sigma \lesssim 0$ nm$^{-2}$.\cite{Podgornik2012} The predicted net negative charge tallies with the mean isoelectric point of $5.0 \pm 1.3$ for a large collection of animal, bacterial and plant viruses.\cite{Graule2010} Available experimental data for the surface charge densities of  $\phi$29 virion, the $\phi29$ prohead, the adenovirus and the minute virus of mice, deduced from atomic force measurements, put $\Delta \sigma$ between about $0.01$ and $0.4$ negative charges per nm$^{-2}$ at near-neutral pH and ionic strengths of 2 and 10 mM.\cite{dePablo2015} For CCMV, electrophoretic mobility measurements, and application of the Henry equation, reveal net surface charge number densities between $\Delta \sigma \simeq -0.1$ nm$^{-2}$ and $\Delta \sigma \simeq +0.03$ nm$^{-2}$ in the pH range from $3$ to $7.5$.\cite{RuizGarcia2014,Gelbart2021}
Finally, entering the bulk solution for $r>R_2$ only mobile ions are present in the solution. The mobile ions in the cavity, for $r<R_1$, and those in the bulk solution, for $r>R_2$, are presumed to be in thermodynamic equilibrium with each other: ions can freely diffuse in and out of the lumen yet we presume that their concentration in the shell region II remains negligible as already mentioned.\cite{Bancroft1984} 

It turns out convenient to work in dimensionless quantities. For that purpose, we define the Debye screening length $\lambda_\mathrm{D} = 1/\sqrt{8\pi\lambda_\mathrm{B}\rho_\mathrm{s}}$ in terms of the Bjerrum length $\lambda_\mathrm{B}$ and the number density of monovalent salt $\rho_\mathrm{s}$. The Bjerrum length is defined as $\lambda_\mathrm{B} = \beta e^2/4\pi\epsilon$ with $\beta=1/k_\mathrm{B} T$ the reciprocal thermal energy, $e$ the elementary charge and $\epsilon$ the dielectric permittivity of the aqueous medium. As usual, $k_\mathrm{B}$ denotes the Boltzmann constant and $T$ the absolute temperature. For room temperature in an aqueous solvent, $\lambda_\mathrm{B} = 0.71$ nm and $\lambda_\mathrm{D} = 0.31/\sqrt{c_\mathrm{s}}$ nm with $c_\mathrm{s}$ the ionic strength of the added monovalent salt in M.
All lengths in our problem shall from this point on be presumed to be scaled by the Debye length, so $r\equiv r/\lambda_\mathrm{D}$, $R_1\equiv R_1/\lambda_\mathrm{D}$ and $R_2\equiv R_2/\lambda_\mathrm{D}$. The electrostatic potential $\psi (r)$ we render dimensionless by defining a Coulomb energy scaled to the thermal energy, so $\psi \equiv \psi e \beta$. For the surface charge density we write $\Delta \sigma \equiv \Delta \sigma \lambda_\mathrm{B} \lambda_\mathrm{D}$, and the magnitude of overcharging we capture in the quantity $\gamma \equiv \Delta \rho / 2\rho_\mathrm{s}$. See also the Introduction.

If we insert the Boltzmann distribution, $\rho_\pm^\mathrm{m} (r)=\rho_\mathrm{s} \exp \left( \mp \psi (r) \right)$, in the dimension-bearing Poisson equation, which reads $\nabla^2 \psi = - e (\rho_+^\mathrm{m}-\rho_-^\mathrm{m}-+\rho_+^\mathrm{i}-\rho_-^\mathrm{i})/\epsilon$, we obtain the following Poisson-Boltzmann equation in dimensionless form
\begin{equation}
    \nabla^2 \psi_\mathrm{I} = \sinh \psi_\mathrm{I} - \gamma
\end{equation}
for the (dimensionless) potential $\psi_\mathrm{I} = \psi$ in region I, so for radial distances $0\leq r < R_1$, 
\begin{equation}
    \nabla^2 \psi_\mathrm{II} = 0
\end{equation}
for the potential $\psi_\mathrm{II}$ in region II where $R_1\leq r < R_2$, and 
\begin{equation}
    \nabla^2 \psi_\mathrm{III} = \sinh \psi_\mathrm{III}
\end{equation}
for the potential $\psi_\mathrm{III}$ region III where $r\geq R_2$, where we refer again to Figure 1. 
These equations have to be supplemented with boundary conditions that enforce overall charge neutrality, radial symmetry and continuity of the potential for all radial positions $r$: i) $\lim_{r\downarrow 0}$ $ \frac{d\psi_\mathrm{I}}{dr} = 0$; ii) $\lim_{r\uparrow R_1} \psi_\mathrm{I} = \lim_{r\downarrow R_1 }\psi_\mathrm{II} $; iii) $ \lim_{r\uparrow R_1} \frac{d\psi_\mathrm{I}}{ dr} = \lim_{r\downarrow R_1} \frac{d\psi_\mathrm{II}}{dr}$; iv) $\lim_{r\uparrow R_2} \psi_\mathrm{II} = \lim_{r\downarrow R_2 }\psi_\mathrm{III}$; v) $\lim_{r\uparrow R_2} \frac{d\psi_\mathrm{II}} {dr} = \lim_{r\downarrow R_2} \frac{d\psi_\mathrm{III}}{dr} +4\pi \Delta \sigma$; vi) $\lim_{r\rightarrow \infty} \psi_\mathrm{III} = 0$. 

We have not been able to solve the above set of differential equations exactly. Hence, we follow the prescription of Lifson,\cite{Lifson1957} who approximately solved the Poisson-Boltzmann equations for the related problem of a model polyelectrolyte chain in which region II is absent as is the surface charge. (A similar approach in a slightly different geometry was used, \textit{e.g.}, by, Ohshima.\cite{Ohshima2008}) Hence, we write $\psi_\mathrm{I} (r) = \phi + \Delta \psi_\mathrm{I} (r)$ for $0\leq r < R_1$, where $\phi$ is a constant background potential and $\Delta \psi_\mathrm{I} (r)$ a position-dependent correction to that constant background potential. If we presume that $|\Delta \psi_\mathrm{I}| \ll |\phi|$, we can Taylor expand $\sinh \psi_\mathrm{I} = \sinh (\phi + \Delta \psi_\mathrm{I}) = \sinh \phi + \cosh \phi \times \Delta \psi_\mathrm{I} (r) + \cdots $. If we insert this in eq.\ (3), and realise that the function $\Delta \psi_\mathrm{I}$ is by construction a varying function of $r$, we obtain the identity $\sinh \phi = \gamma$ for the constant background potential, and
\begin{equation}
    \nabla^2 \Delta \psi_\mathrm{I} = \Gamma^2 \Delta \psi_\mathrm{I},
\end{equation}
for the spatially varying part, where 
\begin{equation}
    \Gamma \equiv \sqrt{\cosh \phi}. 
\end{equation}

It transpires that $\phi$ must indeed be the Donnan potential as we find it to obey eq.\ (1), and is part of the solution to the full, non-linear Poisson-Boltzmann theory in region I. This illustrates once again the deep connection between Donnan theory and Poisson-Boltzmann theory.\cite{Lifson1957,Vrij2013}
The correction $\Delta \psi_\mathrm{I}$ arises in essence because local charge neutrality is broken: mobile ions spread out of the lumen into the bulk solution for reasons of entropy gain. Obviously, the eq.\ (6) that describes this effect can only be accurate under conditions where the linearisation holds. Below we evaluate in more detail the conditions under which that this is the case. 

Since the potential decays significantly in the intermediate region II, we apply the Debye-H\"{u}ckel approximation in region III and linearise eq.\ (5), to obtain
\begin{equation}
    \nabla^2 \psi_\mathrm{III} = \psi_\mathrm{III}.
\end{equation}
By comparing eq.\ (6) with eq.\ (8), we are able to conclude that the Debye length in region I must be a factor $\Gamma =\sqrt{\cosh \phi} \geq 1$ \textit{smaller} than that in the outer region III.\cite{Lifson1957} This enhanced electrostatic screening is stronger the larger the magnitude of the Donnan potential. Indeed, for Donnan potentials stronger than the thermal energy, so for $|\phi| \gtrsim 1$, we have $\Gamma \sim \sqrt{|\gamma|}$, implying that in that case the effective Debye length in region I, $\lambda_\mathrm{D}/\Gamma$, scales as $1/\sqrt{|\Delta \rho|}$. The magnitude of this effective Debye length is then determined by the mismatch between the number of immobile positive and negative charges in the lumen, and must then be virtually independent of the concentration of salt in the bulk solution.

The remaining set of equations, eqs.\ (4), (6) and (8) can now straightforwardly be solved in polar co-ordinates and using the boundary conditions quoted above in order to fix all the integration constants. For the dimensionless potential in region I, we find
\begin{equation}
    \psi_\mathrm{I} (r) =\phi + \Delta \psi_\mathrm{I} = \phi - u \times \left( \frac{\sinh r\Gamma}{r\Gamma} \right),
\end{equation}
with
\begin{equation}
    u  \equiv \frac{\phi (1+R_2) - 4\pi\Delta \sigma R_2}{\left( 1+R_2-R_1 \right)\cosh (R_1\Gamma) + \Gamma^{-1} \sinh (R_1\Gamma)},
\end{equation}
a parameter that depends on the inner and outer radii of the protein shell $R_1$ and $R_2$, the degree of overcharging via the Donnan potential $\phi$ and the net surface charge density $\Delta \sigma$. Clearly, if $|u\sinh (R_1\Gamma) / R_1\Gamma| \ll |\phi|$, then $\psi_\mathrm{I} \sim \phi $ for $ r<R_1$, and the Donnan potential provides a good representation of the electrical potential in region I, the lumen of the protein shell. 

The potential in the protein shell (region II) obeys
\begin{equation}
    \psi_\mathrm{II} (r) = \psi_\mathrm{I}(R_1) - u \times \left(\cosh R_1\Gamma - \frac{\sinh R_1\Gamma}{R_1\Gamma} \right) \times \left(1 - \frac{R_1}{r}\right),
\end{equation}
which is not uniform, unless $u=0$. This happens exactly if $\phi = 4\pi \Delta \sigma R_2/(1+R_2)$, in which case the uniform Donnan potential extends all the way to the boundary with region III, the outer surface of the shell. The potential in 
region III decays with increasing radial distance $r$ under all circumstances, and reads
\begin{equation}
    \psi_\mathrm{III} (r) = \psi_\mathrm{II}(R_2) \times  \frac{R_2}{r} \exp \left( -r + R_2 \right).
\end{equation}
The somewhat unwieldy expressions eq.\ (9) -- (12) agree with those of Lifson for the corresponding case $R_1=R_2$ and $\Delta \sigma = 0$.\cite{Lifson1957} Lifson found excellent agreement with a numerical solution of the equations for all cases investigated, that is, for fixed $\gamma = 5$ and varying $R_1 > 1$, and for fixed $R_1=5$ and varying $\gamma \geq 0.5$. 

To calculate the local pH and relate that to the value in the bulk solution for $r\rightarrow \infty$, we again make use of the Boltzmann distribution. Let $[\mathrm{H}^+](r)$ be local concentration of H$^+$ ions, and $[\mathrm{H}^+]_\mathrm{s}$ that of the bulk solution. The local concentration of $\mathrm{H}^+$ ions depends on the local potential $\psi$ via the relation $[\mathrm{H}^+](r) = [\mathrm{H}^+]_\mathrm{s} \exp \left(- \psi(r) \right)$. Hence, the pH 
depends on the radial co-ordinate and becomes only equal to the bulk value if $\psi \rightarrow 0$, so for $r\rightarrow \infty$.
Focusing attention on the interior of the capsid,  region I, we must have a local concentration of  $[\mathrm{H}^+] = [\mathrm{H}^+]_\mathrm{s} \exp \left(- \psi_\mathrm{I}(r) \right)$. 
The \textit{average} concentration of H$^+$ ions in region I, $[\mathrm{H}^+]_\mathrm{av}$, is equal to
\begin{equation}
    [H^+]_\mathrm{av}/[H^+]_\mathrm{s} \equiv \frac{3}{R_1^3} \int_0^{R_1} dr r^2 \exp \left( - \psi_\mathrm{I} \right),
\end{equation}
in terms of the scaled variables $r$, $R_1$ and $\psi_\mathrm{I}$. Note that eq.\ (13) applies to any positively charged mobile ionic species, whilst for negatively charged ionic species we only need to replace the minus sign in the exponential by a plus sign. Finally, the pH differential between the interior of the capsid and the bulk solution is given by 
\begin{equation}
    \Delta \mathrm{pH} \equiv - \log_{10} [H^+]_\mathrm{av}/[H^+]_\mathrm{s},
\end{equation}
which we can calculate once the potential in cavity is known, $\psi_\mathrm{I}$. Notice that for $|u| \rightarrow 0$, we have $J=1$ and we retrieve eq.\ (2) noting that $\log_{10} \textrm{e} = 1/ \ln 10$ with e Euler's number.

To calculate the pH shift, $\Delta \mathrm{pH}$, we insert eq.\ (9) into eq.\ (13), and by applying a suitable change of variables we find from eq.\ (14)
\begin{equation}
     \Delta \mathrm{pH} = \frac{\phi -\ln J}{\ln 10} 
\end{equation}
with
\begin{equation}
    J = \frac{3}{R_1^3\Gamma^3} \int_0^{R_1\Gamma} ds s^2 \exp\left( u \times \frac{\sinh s}{s}\right).
\end{equation}
We have not been able to exactly solve this integral. We expect that under most experimentally relevant conditions $|u \sinh (R_1\Gamma) /R_1 \Gamma | \lesssim 1$, allowing us to Taylor expand the exponential and obtain to second order in the parameter $u$,
\begin{equation}
    J = 1 + 3u \left(\frac{R_1\Gamma \cosh R_1\Gamma - \sinh R_1\Gamma }{R_1^3 \Gamma^3 }\right) + 3u^2  \left(\frac{\cosh R_1\Gamma \sinh R_1\Gamma -R_1\Gamma}{2R_1^3\Gamma^3}\right)+ \cdots.
\end{equation}
In the limit $R_1 \Gamma \ll 1$, the corrections in powers of $u$ are small only if $|\phi - 4\pi\Delta \sigma R_2|\ll1$. For $R_1\Gamma \gtrsim 1$, they are of the order 
$(\phi - 4\pi \Delta \sigma)/(R_1 \Gamma)^2$ implying that as long as the radius of the lumen is much larger than the effective screening length in it, the pH shift should be dominated by the Donnan potential. 
In the opposite limit, $|u \sinh ( R_1\Gamma) /R_1 \Gamma| \gg 1 $, the perturbation approach that we invoke to solve the Poisson-Boltzmann equation in region I should break down. The reason is that in that case $|\Delta \psi_\mathrm{I}|$ is not necessarily small compared to $|\phi|$ near the edge of the lumen. Hence, we do not discuss this limit any further. 

\section{Results}
To investigate in more detail the limits of applicability of Donnan theory, we compare its predictions with our perturbation theory. From equation (9) we read off that provided $|u\sinh (R_1\Gamma) / R_1\Gamma | \ll |\phi|$ we have $\psi_\mathrm{I} (r) \simeq \phi$ for all $0\leq r\leq R_1$, implying that in that case the Donnan potential accurately describes the electrical potential in the lumen of the capsid shell. This happens if at least one of two conditions is met:
\begin{itemize}
    \item[i)] The presence of a net charge on the outer surface of the shell compensates for the drop in the potential in the shell, so if $\phi = \sinh^{-1} \gamma \simeq 4\pi \Delta \sigma R_2/(1+R_2)$. In that case, we have $\psi_\mathrm{I} \simeq \psi_\mathrm{II} \simeq \phi$, and most of the drop of the potential happens outside of the shell, that is, in region III. This shows that a net surface charge, if not too large and of the same sign as the net immobile space charge present in the lumen, makes the prediction of Donnan theory more accurate than without it;
    \item[ii)] If the concentration of salt or the magnitude of the Donnan potential is sufficiently large so that the effective screening length of the solution in the cavity is much smaller than the radius of the lumen, that is, if $R_1\Gamma = R_1 \sqrt{\cosh \phi} = R_1 \sqrt{\cosh (\sinh^{-1} \gamma)} \gg 1$. For weak degrees of over- or undercharging $|\gamma|\ll 1$, this implies that $R_1 \gg 1$, whilst for large degrees of overcharging $|\gamma|\gg 1$, the dimensionless radius $R_1 \gg 1/\sqrt{\gamma}$ may actually be substantially smaller than unity.
\end{itemize} 
From this we conclude that in the present context Donnan theory has a wider range of applicability than is sometimes thought. It is neither restricted to low Donnan potentials\cite{Grodzinsky1993} nor to low ionic strengths\cite{Huster1999}, as in fact is already clear from the early work of Lifson.\cite{Lifson1957} It seems that modelling a continuous immobile space charge distribution by a series of fixed, localised charges, as was done by Grodzinsky \textit{et al.\ }\cite{Grodzinsky1993} and Huster \textit{et al.\ }\cite{Huster1999}, would lead us to underestimate the range of applicability of Donnan theory.

\begin{figure}[ht!]
\centering
\includegraphics[scale=0.6]{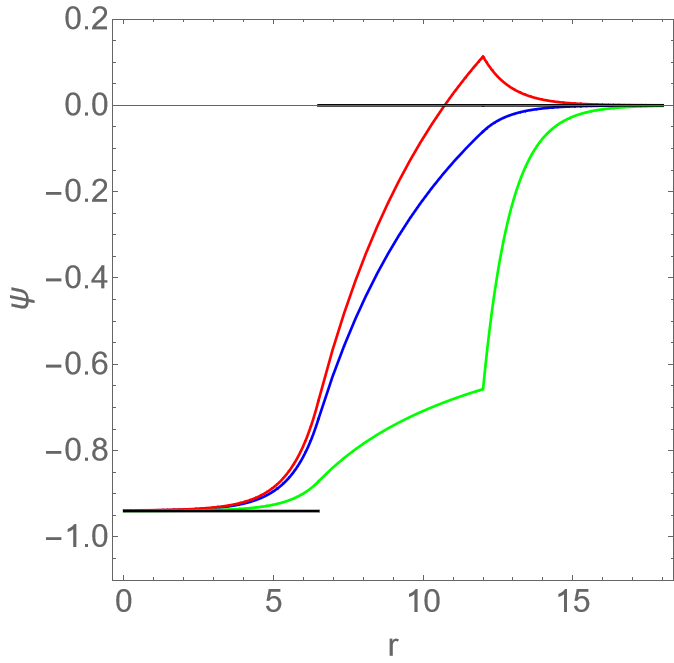}
\caption{Dimensionless potential $\psi$ as a function of the dimensionless radial distance $r$ from the centre of a $T=1$ virus-like particle under conditions of 0.15 M ionic strength. The dimensionless radius of the lumen is $R_1 =6.5$ and that of the shell $R_2=12$. From bottom to top: predictions from Poisson-Boltzmann theory with dimensionless surface charge densities $\Delta \sigma= -0.055, 0, +0.016$. The step function represents the Donnan potential $\phi=-0.94$ that gives the corresponding average pH difference between lumen and bulk of $-0.4$.\cite{Cornelissen2018} See also the main text.}
\end{figure}

We now illustrate our findings by taking as model parameters those that we estimate for the $T=1$ virus-like particles investigated by Maassen and collaborators.\cite{Cornelissen2018} We recall that these are formed by the spontaneous co-assembly of coat proteins of CCMV and poly(styrene sulfonate) copolymers. We set the dimensionless shell radii equal to $R_1 =6.5$ and $R_2=12$, given the experimental ionic strength of 0.15 M and our estimates for the inner and outer radii of the shell for which we take 5 nm and 9.5 nm, respectively. The total number of positive charges on the RNA binding domains equals $+600$. For pHs in the range from 3 and 7.5, around the isoelectric point $\mathrm{pI}\simeq 4$ of the wild-type CCMV virus particles, electrophoretic mobility measurements reveal net surface charge number densities between $\Delta \sigma \simeq -0.1$ nm$^{-2}$ and $\Delta \sigma \simeq +0.03$ nm$^{-2}$.\cite{RuizGarcia2014} From this, we conclude that the corresponding dimensionless values $\Delta \sigma$ must be in the range from $-0.055$ to $+0.016$. To produce a pH shift of $-0.4$ at pHs larger than 6 found experimentally, we obtain from eq.\ (15) a value of $\phi=-0.94$ if we set $\Delta \sigma=-0.055$ that should be accurate for pHs larger than 6. This gives for the degree of overcharging $\gamma = -1.08$. These findings are very close to the values of $\phi=-0.92$ and $\gamma=-1.06$ that we obtain from the Donnan theory for the same system. 

Figure 2 shows the dimensionless potential $\psi$ as a function of the dimensionless radial co-ordinate $r$ for the dimensionless surface charge densities $\Delta \sigma = -0.055, 0, +0.016$. Notice the discontinuity in the radial derivative of the potential for $r=R_2$ for $\Delta \sigma \neq 0$. Also indicated is the Donnan potential, which is non-zero only in region I, for $r<R_1 = 6.5$. The figure confirms that the Donnan potential reasonably accurately describes the potential in the cavity of the shell for the range of surface charge densities indicated. 
Only very near the inner surface of the shell the Donnan potential and the potential obtained from Poisson-Boltzmann theory deviate from each other. The discrepancy is particularly small (under ten per cent) if the outer surface is negatively charged. For the vast majority of viruses this seems to be the case under conditions of neutral pH.\cite{Graule2010} In view of the above, the agreement is not entirely unexpected, of course, as for the solution conditions of the experiments of Maassen \textit{et al.} the value of $R_1\Gamma \simeq 7.8 $ is quite larger than unity.\cite{Cornelissen2018} 

\begin{figure}[ht!]
\centering
\includegraphics[scale=0.6]{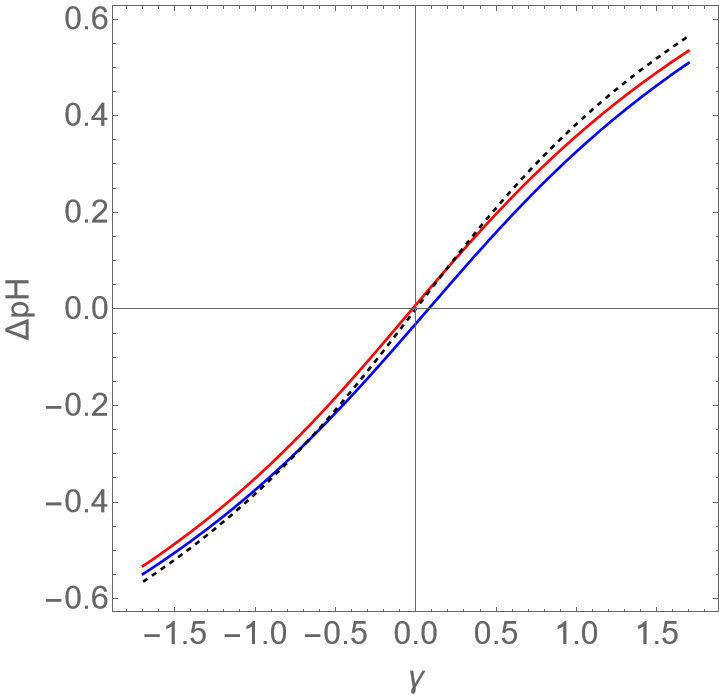}
\caption{Difference $\Delta$pH between the average pH in the lumen of a $T=1$ virus-like particle and the pH of the bulk solution as a function of a measure of the imbalance of the number of immobile positive and negative space charges $\gamma$ in the lumen of the capsid. The dimensionless radius of the lumen is $R_1 =6.5$ and that of the shell $R_2=12$ at the fixed ionic strength of 0.15 M. Dashed curve: prediction from Donnan theory. Drawn curves: $\Delta \sigma = -0.055$ (in blue, bottom) and $\Delta \sigma = +0.016$ (in red, top). } 
\end{figure}

Figure 3 shows the predicted pH shift $\Delta$pH as a function of the degree of over- or undercharging $\gamma$ for dimensionless surface charge densities of $\Delta \sigma =-0.055$ and $+0.016$, and compares these with the predictions of Donnan theory for the $T=1$ virus-like particles with $R_1 =6.5$ and $R_2=12$ for the fixed ionic strength of 0.15 M, mimicking the experiments of Maassen and collaborators\cite{Cornelissen2018} but now allowing for a variable overcharging. Agreement between our Poisson-Boltzmann theory and Donnan theory is indeed rather good. The figure confirms that the accuracy of Donnan theory increases if the net surface charge has the same sign as the net immobile space charge in the lumen. Finally, Donnan theory slightly overestimates $|\Delta$pH$|$ for larger positive or negative values of $\gamma$. This we attribute to the Debye-H\"{u}ckel approximation becoming less accurate in the outer region III for relatively large degrees of over- or undercharging. 

The question arises what happens to the usefulness of Donnan theory if the ionic strength were a factor of, say, ten lower, so 15 mM instead of 150 mM. We would then have $R_1 \simeq 2.1$ and $R_2 \simeq 3.8$, which could indicate that Donnan theory might in that case be not quite as accurate. However, if we assume the overcharging to remain more or less constant, then the degree of overcharging becomes much more negative with $\gamma = -11$. In that case, we find $\phi = -3.1$ and $\Gamma = 3.3$. This means that for a ten times smaller ionic strength we have $R_1\Gamma = 6.9$, implying that Donnan theory should remain to be reasonable accurate. Indeed, the pH shift predicted by Donnan theory amounts to $-1.3$ and that from our Poisson-Boltzmann theory is also $-1.3$, presuming a dimensionless surface charge density of $-0.055$. We conclude that an increase in the bulk screening length, which in principle would make the Donnan theory less accurate, is more than compensated for by an increase in the magnitude of the Donnan potential that decreases the effective Debye length in the lumen of the virus-like particles.

\begin{figure}[ht!]
\centering
\includegraphics[scale=0.6]{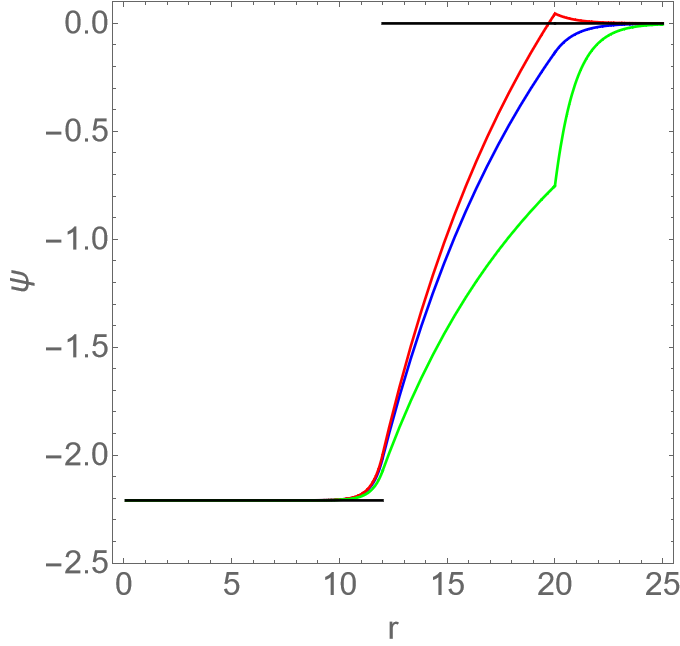}
\caption{Dimensionless potential $\psi$ as a function of the dimensionless radial distance $r$ from the centre of a $T=3$ virus-like particle under conditions of 0.16 M ionic strength, 2880 positive charges on the RNA binding domains and 5400 negative charges on the encapsulated polyanion mimicking the experiments of Ren and collaborators.\cite{Lim2006} For the dimensionless radius of the lumen we set $R_1 =12$ and for that of the shell $R_2=20$. From bottom to top predictions from Poisson-Boltzmann theory with dimensionless surface charge densities $\Delta \sigma= -0.055, 0, +0.016$. The step function represents the Donnan potential $\phi$. See also the main text.}
\end{figure}

Let us now apply our theory to the experiments of Ren and collaborators, who encapsulated poly(styrene sulfonic acid) PSA of varying molecular weights from 13 to 990 kDa  using the coat proteins of horseradish chlorotic ringspot virus or HCRSV, and \textrm{predict} the pH shift.\cite{Lim2006} PSA is a strong polyacid with a pK$_\textrm{a} \simeq 1$, and is fully charged at the pH of about five at which the experiments were done. Ren and collaborators found a more or less constant loading of PSA irrespective of the molecular weight of the PSA, corresponding to a single molecule of 990 kDa. This translates to an encapsulation of 5400 negative charges. The total number of positive charges on the arginine-rich motifs of the $T=3$ virus amounts to 2880, suggesting an overcharging of about a hundred per cent.\cite{Dokland2003} At an ionic strength of 0.16 M, and taking for the radius of the lumen 9 nm and that of the shell 15 nm, we get for the scaled dimensions of the shell $R_1=12$ and $R_2=20$. 

Hence, our measure for the degree of overcharging attains the value of $\gamma = -4.3$, the corresponding Donnan potential $\phi = -2.2$ and the reduction factor of the Debye length $\Gamma = 2.1$. We conclude that Donnan theory should be rather accurate, as Figure 4 in fact confirms for $-0.055 \leq \Delta \sigma \leq +0.016$. Notice that the so-called zeta potential of the virus-like particles measured by Ren \textit{et al.} has values between between $-2.3$ and $-2.9$ mV,\cite{Lim2006} which translates to a net surface charge number density of about $-0.016$ nm$^{-2}$ if we apply the Henry equation.\cite{Hunter1981} Application of the Henry equation is justified, provided capsids are impermeable to the flow field. This seems to be the case for capsids of simple plant viruses\cite{Gelbart2021} albeit that it need not be generally true for all types of particle.\cite{Duval2008,Duval2022} Presuming the quoted value for the surface charge density to be accurate, the corresponding dimensionless surface charge density becomes $\Delta \sigma \simeq - 8.5 \cdot 10^{-3}$ for the given ionic strength of the solution. The pH shifts we then obtain from Donnan theory and from Poisson-Boltzmann theory are for all intents and purposes equal, namely $\Delta \mathrm{pH} = -0.94$.

\section{ Discussion and Conclusions}
An imbalance in the number of localised positive and negative space charges in the lumina of viruses, virus-like particles and other types of protein shell, semi-permeable to mobile ionic species, gives rise to a Donnan potential difference between the bulk solution and the inside of the particles. This potential difference induces a pH differential that can potentially be large. This, in itself, is not surprising.\cite{Donnan1924} What is surprising, given the small size of the particles, is that simple Donnan theory, is remarkably accurate in predicting the magnitude of this pH shift when compared with predictions from Poisson-Boltzmann theory, as it presumes local charge neutrality and homogeneous distributions of mobile ions in- and outside the shell. Under typical experimental conditions, the presence of a net charge on the outer surface of the shell turns out to have a relatively minor effect. For overcharged virus-like particles the presence of a negative surface charge actually makes Donnan theory more rather than less accurate. 

The reason why Donnan theory works so well is that, in practise, the Donnan potential in the lumina of the particles is sufficiently large so that the effective (local) screening length in it becomes small on the scale of the width of the lumen the region in it where most of the immobile charges can be found.  A similar conclusion can be drawn for other types of particle characterized by an immobile space charge distribution.\cite{Lifson1957,Ohshima2008,Duval2015} Consequently, even for the smallest of virus-like particles the concentration of mobile ions remains more or less constant in the 10 nm or so wide lumen, at least if the immobile space charge distribution is also more or less uniform. Even if the immobile charges are not uniformly distributed in the lumen but localised in a region near the inner surface of the shell, we would still expect the predictions for the pH shift in that region to be well-described by Donnan theory as long as it is (much) wider than the effective screening length.\cite{Duval2008} 

All of this implies that simple Donnan theory can indeed be used to estimate the total amount of encapsulated material inside of small proteinaceous shells from the measured pH, as was done by Maassen \textit{et al.} for virus-like particles self-assembled from the coat proteins of the plant virus CCMV and a polyanionic cargo.\cite{Cornelissen2018} Conversely, if the amount of encapsulated polyanionic material is known, such as in the experiments of Gelbart \textit{et al.}\cite{Gelbart2008,Gelbart2011} and of Lim \textit{et al.}\cite{Lim2006}, Donnan theory can be used to estimate the pH in the lumen of the virus-like particle. As discussed in the Introduction, the predicted pH shifts can be as large as a full pH unit,  in both positive and negative directions. If virus-like particles are used for catalysis, e.g., by way of encapsulated enzymes, then the activity of these enzymes may be influenced not only by the compartmentalisation itself but also by the local pH and ionic strength.  \cite{Cornelissen2011,vanHest2016} Since the charged state of enzymes depends on the pH, they would modify the Donnan potential self-consistently if their number is large enough. In fact, the coat protein itself might be involved in buffering activity.\cite{Cornelissen2018}
This kind of charge regulation can be incorporated in both Donnan and Poisson-Boltzmann theory relatively straightforwardly.\cite{Mohwald2007,Duval2008,Podgornik2022} 

For instance, if the cargo is a weak polyacid, such as polyacrylic acid,\cite{Lim2006} we only need to modify the parameter $\gamma = (\rho_+^i - \rho_-^i)/2\rho_\mathrm{s}$, which now becomes a function of the local potential $\psi_\mathrm{I}$ in the lumen, for
\begin{equation}
   \frac{\rho_-^i}{\rho_{-,0}^i} = \frac{1}{1+10^{-\mathrm{pH_{in}} + \mathrm{pK_a}  }} = \frac{1}{1+10^{-\mathrm{pH_{out}} + \mathrm{pK_a} - \frac{\psi_\mathrm{I}}{\ln 10}}}.
\end{equation}
This, of course, is the familiar Henderson-Hasselbalch equation, where in the second equality we have expressed the local pH in the lumen, $\mathrm{pH_{in}}$, in terms of the pH in the bulk solution, $\mathrm{pH_{out}}$. Further, $\mathrm{pK_a}$ is the dissociation constant of the weak polyacid, and $\rho_{-,0}^i$ the concentration of chargeable groups on the encapsulated polyacid that we again presume to be uniformly distributed in the lumen. At the level of Donnan theory, $\psi_\mathrm{I} = \phi$, whilst within our Poisson-Boltzmann theory, we would write again $\psi_\mathrm{I} = \phi + \Delta \psi_\mathrm{I}$, and expand $\sinh \psi_\mathrm{I}$ as well as $\gamma$ to first order in $\Delta \psi_\mathrm{I}$. The Debye length renormalisation factor $\Gamma$ now obeys the equality $\Gamma^2 = \cosh \phi - \partial \gamma/\partial \psi_\mathrm{I}|_{\psi_\mathrm{I} = \phi}$. As long as $R_1\Gamma$ remains sufficiently large, Donnan theory should again be reasonably accurate. 

Eq.\ (18) tells us that the pH shift $\Delta \mathrm{pH}$ now depends on the pH of the bulk solution $\mathrm{pH_{out}}$, except if $|\mathrm{pH_{in}} - \mathrm{pK_a}| \gtrsim 1$. For $|\mathrm{pH_{in}} - \mathrm{pK_a}| \lesssim 1$, the pH shift can actually compensate for changes in the outside pH, leading to a constant pH inside the particle. This happens if $\partial \Delta \mathrm{pH} /\partial \mathrm{pH_\mathrm{out}} = -1$, which within Donnan theory translates to  $\partial \phi / \partial \mathrm{pH_{out}} = - \ln 10$. It shows that charge regulation can indeed lead to a buffering effect, explaining the findings of Maassen and collaborators, who find a constant (negative) pH shift for solution pHs from 6 to 8 but a constant inside pH for solution pHs between 5 and 6.\cite{Cornelissen2018} In their particular case the buffering is arguably not caused by the encapsulated strong polyanion but by moieties on the coat protein itself.

We shall not dwell on this issue any further, and end by mentioning that the calculation involving charge regulation would allow us to establish the mean charge on any encapsulated weak polyacid and compare that with the mean charge of the same polyacid in free solution. This, with the aforementioned experiments of Lim and collaborators in mind.\cite{Lim2006} Note that charged state of the weak polyacids in free solution and that in the capsids are in principle not the same due to the impact of what essentially is the Donnan potential.\cite{vanderSchoot2015} We leave this for future work.

\begin{acknowledgement}
The authors thank Profs. Albert Philipse (Utrecht University) and Rudolf Podgornik (University of Ljubljana) for discussions and helpful suggestions. I am also grateful to dr. Stan Maassen for providing the table-of-contents image. This research is supported by the Eindhoven University of Technology.

\end{acknowledgement}

\bibliography{Donnan}

\end{document}